\documentclass[preprint,superscriptaddress]{revtex4}
\usepackage{graphicx}
\usepackage{amsmath, amsthm, amsfonts,amssymb}

\begin{document}
\title{Dynamics of clustered opinions in complex networks}
\author{Woo-Sung Jung}
\email{wsjung@physics.bu.edu}
\affiliation{Center for Polymer Studies and Department of Physics, Boston University, Boston, MA 02215, USA }
\affiliation{Department of Physics, Korea Advanced Institute of Science and Technology, Daejeon 305-701, Republic of Korea}

\author{Hie-Tae Moon}
\affiliation{Department of Physics, Korea Advanced Institute of Science and Technology, Daejeon 305-701, Republic of Korea}

\author{H. Eugene Stanley}
\affiliation{Center for Polymer Studies and Department of Physics, Boston University, Boston, MA 02215, USA }               

\begin{abstract}
A simple model for simulating tug of war game as varying the player number in a team is discussed to identify the slow pace of fast change. This model shows that a large number of information sources leads slow change for the system. Also, we introduce an opinion diffusion model including the effect of a high degree of clustering. This model shows that the \textit{de facto standard} and \textit{lock-in effect}, well-known phenomena in economics and business management, can be explained by the network clusters. 
\end{abstract}

\maketitle
\section{Introduction}
Network models are an active research field in recent, and  many stylized features, such as a short average length between nodes, a high clustering coefficient and a power-law distribution, have been detected in many real network systems in nature and society \cite{watts99,barabasi02,amaral00,watts98,albert99,albert02,klemm02,klemm02b,klemm03,vazquez03}. The dynamics in small-world and scale-free networks is one of the recent issues to be addressed in the study of complex networks. A small-world network \cite{watts99,watts98} is generated by rewiring the links, and the scale-free network introduced by Barab\'asi \textit{et al.} \cite{barabasi02,albert02} shows a power-law connectivity distribution. Recently, Klemm \textit{et al.} \cite{klemm02,klemm02b} introduced an algorithm for a highly clustered scale-free network model. In particular, these network models are used to investigate the dynamics of interaction elements in physics, biology, economics, business management, and many other areas.

Econophysics is one of the most active interdisciplinary fields \cite{mante00,arthur97,mante99,yamasaki05,jung06,eguiluz00,krawiecki02,yang06,chowdhury99,cont00}. One of the topics most widely studied at present is complex network models because real financial markets have many interacting agents with a huge amount of information. In addition, microscopic models including the spin model describe well the properties of financial markets \cite{eguiluz00,krawiecki02,yang06,chowdhury99,cont00}. The agents are represented by spins and the interaction between agents by fields in the model. This microscopic model is also widely applied to social systems. In fact, economic and social systems are not very different from the viewpoint of complex networks \cite{valla94,stauffer05,sznajd00,latane81,nowak90,holyst00,suchecki05,breban05,lind06,jo06}. There are many interactions and communications between agents in the network, and these are subject to opinion dynamics. Thus, complex network theory is a powerful instrument in interdisciplinary areas such as econophysics. Agent-based models and game theory are useful approaches for investigating economic and social networks \cite{axelrod84,myerson97,gibbons92,miller03,challet97,arthur94,johnson99,marsili01,lee06}. Agent-based models were first considered in the late 1940s. In this type of model, agents with certain properties are assumed and then simulations are carried out to model real phenomena. In this context, we investigate opinion diffusion in complex networks using agent-based model.

\section{Tug of war game}
In this section, we introduce a simple game which motivates the study of the following section. Tug of war game, a simple test of strength which was part of the Olympic Games from 1900 to 1920, is investigated using computer simulation. The game rule is very simple; two teams align themselves at opposite ends of a rope and pull against each other. In our simulation, the player number in a team is varied from 10 through 10000. Every player's original stamina, $s_i(t=0)$, is 1.0, and it decays by his or her particular \textit{stamina decay rate} $\gamma_i$ determined randomly between 0.9 to 1.0. In other words, a player's stamina is derived as $s_i(t+1)=\gamma_i s_i(t)$. Team strength is simply defined as the sum of the stamina of all the players in the team. When the strength of one team exceeds that of the other team by 10\%, the winner is determinated and the game is over. If the strength of the two teams is balanced for a long time, the players become exhausted and the game ends in a draw. Our game is played within 1000 time steps.

\begin{figure}
\begin{center}
\resizebox{0.7\textwidth}{!}{
\includegraphics{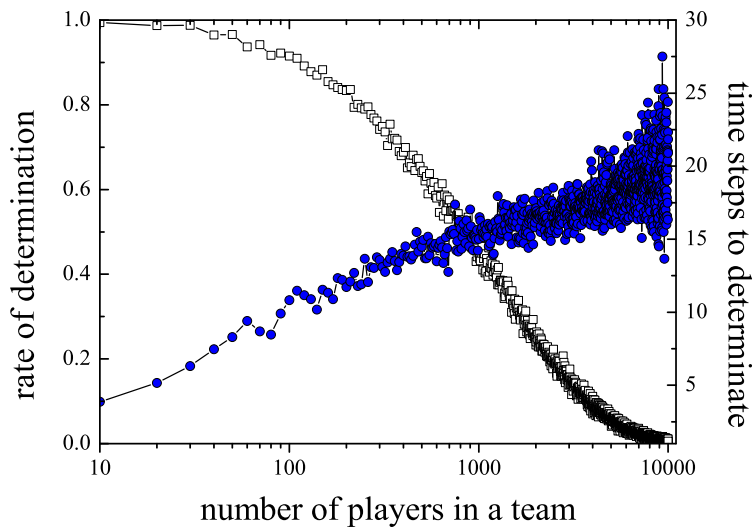}
}
\caption{Result of the \textit{tug of war} game. Squares  show the rate of determination, and circles the number of time steps required to determinate the winner.} \label{tugofwar}
\end{center}
\end{figure}

Fig. \ref{tugofwar} shows the result of our tug of war game. In this figure, circles show \textit{the number of time steps} required to determinate the winner. However, every game does not have the winner because some games end in a draw from the long balance of the power. We consider only the games having the winner to calculate the number of time steps (circles), and define \textit{the rate of determination} (squares in Fig. \ref{tugofwar}) as the fraction of the game having the winner over the whole game. For instance, we tested 1000 games between two teams of 10 players, and got the result that 994 games can determinate the winner. The rate of determination for this case is 0.994. As the number of players increases, the determination of the winner becomes more difficult. Even if there is a weak player in one team, the stamina of the other players can make up for this weak point. These complements are abundant with a large number of players in the team, and may lead to a greater number of drawn games. When the game is played between two teams of 10 players each, the winner can be determinated after three time steps in almost all games. However, in a game with two teams of $10000$ players, most games finish before the winner is determinated because players are totally exhausted from the long power struggle between the two teams. Even though the winner may be determinated by chance, this requires more than 20 time steps.  A large number of players in a team means that we have more factors (agents in this game) to consider and need more time to calculate including these factors.

This is an example of \textit{the slow pace of fast change} \cite{chakra2003}. Nowadays, the market occasionally accepts innovations rather slowly compared to the superior technological advances of the innovation. The market is currently huge and is more complex than ever before. The amount of information has considerably increased and an agent has a large amount of information to consider. It is not hard to imagine that an agent spends much time thinking and analyzing before making a decision. Therefore, a change in personal decision requires more time than before. The large number of player number in tug of war game corresponds to the abundant information. The determination of the winner is directly connected to the decision making.

A node of complex network has many neighbors including nearest, second-nearest, third-nearest ones, and so on, and the average path length of complex network is short. In the study of opinion dynamics, an agent usually interacts with its neighbors. Neighbors give the information to the agent, and the opinion diffuses quickly due to the short path length. Therefore, when we investigate the opinion diffusion on complex network, the whole network usually has only one opinion in a few time steps.

However, real world consists of various opinion groups and the various opinions do not easily converge into one. It is necessary to consider more features for studying opinion dynamics. We focus on clusters of complex network. If all neighbors of a given agent have the same information, the agent will gather only one type of information and will not change his/her mind in the future. Finally, the agent and the neighbors will form a cluster on this basis. If the network consists of several clusters, the whole network is prevented from having only a single opinion. The result of this section is a motivation of the following one, the opinion diffusion model with clusters on complex network.

\section{Opinion diffusion and marketing strategy}
In this section, we investigate additional model rahter different from that of the previous section but related to each other. The $N$ agents in a given network have particular orientations $\sigma_i(t)=\pm 1$ at discrete time step $t$. Usually, this spin model is used to study the financial market by physicists as the orientation corresponds to the opinion to sell ($-1$) or buy ($+1$) \cite{eguiluz00,krawiecki02,yang06,chowdhury99,cont00}. However, the application of this spin model can be expanded to social opinion models of binary states such as yes ($+1$) or no ($-1$). From now, we describe this model as a study of the financial market for the convenience of description. The orientation of agent $i$ depends on the opinion of his neighbors as follows:

\begin{equation}
\sigma_{i} (t+1) = \left\{
\begin{array}{ll}
+1&\textrm{with probability }p,\\
-1&\textrm{with probability }1-p,
\end{array} \right.
\label{eq:sigma}
\end{equation}
where
\begin{equation}
p=\frac{1}{1+e^{-2 I_i(t)}} .
\label{eq:p}
\end{equation}

The local field, $I_i (t)$, is the sum of the orientation of the neighbors and is defined in previous studies \cite{krawiecki02,yang06} as:
\begin{equation}
I_i(t)=\frac{1}{M}\sum_j \sigma_j (t),
\label{M}
\end{equation}
where $j$ means the neighbors of agent $i$ and $M$ is the number of those neighbors. The log return (or price change) of the model at time $t$ is
\begin{equation}
x(t)=\frac{1}{N} \sum \sigma_i (t),
\end{equation}
and $x(t+1)$ is simply derived to $\tanh x(t)$. Under this assumption, $x(t+1)$ is derived as follows:
\begin{equation}
x(t+1)=\frac{1}{N} \sum \sigma_i (t+1)=2p-1=\frac{e^I -e^{-I}}{e^I+e^{-I}}=\tanh I_i (t) = \tanh x(t),
\end{equation}
and $x(t)$ goes to zero as the time flows. However, we redefine the local field as
\begin{equation}
I_i(t)=\sum_j \sigma_j (t),
\label{noM}
\end{equation}
which makes $x(t)$ converge to +1 or -1 because $x(t+1)=\tanh Mx(t)$, and use it because we  investigate famous phenomena, \textit{de facto standard} and \textit{lock-in}, and the convergence to +1 or -1 is more suitable to explain them. 

First, we apply this model with 1000 agents to the Watts--Strogatz small-world network model (the WS model) \cite{watts98}. However, all agents in the network have the same opinion and the log return $x_i(t)$ rapidly approaches $+1$ or $-1$. The long-range interactions of the network makes opinion diffusion very fast and saturation occurs after a few iterations. We also investigate the Barab\'asi--Albert scale-free network model (the BA model) \cite{albert99,albert02}, but the result is similar to that of the WS model. 

Recently, an algorithm for the highly clustered scale-free network model (the structured model), which has several high clusters, was introduced by Klemm \textit{et al.} \cite{klemm02}. When we apply our model to this network model, the result is different from those of the WS and BA models. We find a struggle for power between optimists and pessimists due to the clusters of the network (Fig. {\ref{fig:bavshcsf}). Klemm \textit{et al.} also analyzed the features of self-organizing networks with scale-free and small-world behaviors as a crossover between the structured model and the BA model, adjusting the parameter $\mu$ (the growing model) \cite{klemm02b}. The structured model is constructed for $\mu=0$, and the BA model for $\mu=1$. We show the result with the $\mu$ value of $10^{-3}$ as a representative of the growing model in Fig. \ref{fig:bavshcsf}.

\begin{figure}
\begin{center}
\resizebox{0.7\textwidth}{!}{
\includegraphics{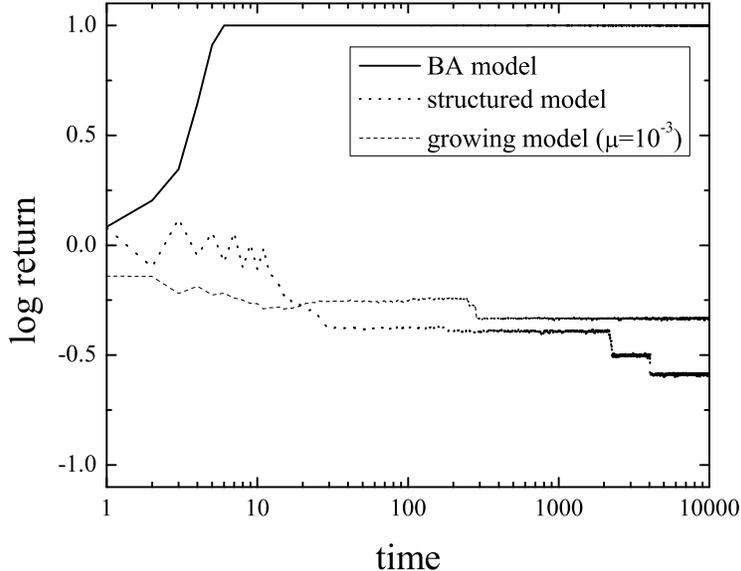}
}
\caption{Log returns in (a) the Barab\'asi--Albert scale-free network model (upper solid-line) and (b) the  structured scale-free network model (lower dot-line). The middle dash-line is log return of (c) the growing scale-free network model when the $\mu$ is $10^{-3}$.}
\label{fig:bavshcsf}
\end{center}
\end{figure}

The result is sensitive to the initial condition. However, the log return of the BA model always approaches +1 or -1 in a few time steps because the majorities who have one opinion are formed very rapidly. In the case of the structured model, the log return reaches a certain value that is different from +1 and -1 because of the struggle between several clusters consisting of opinion +1 or -1. It also approaches +1 or -1 at the end, but very slowly than that of the BA model. In addition, sometimes a few agents or clusters change their opinion to the opposite, which leads to jumps in the log return (Fig. \ref{fig:bavshcsf}). 

The structured model is basically one-dimension network, which has similar properties with one-dimension lattice. The number of neighbors who have interactions are small but the localities strong in the lattice, and the ordered state of the whole system is not appeared. 
Thus, the result of the structured model can be from the one-dimensional structure than the high clusters. To investigate this more detail, we tested our model in the growing model as varying $\mu$ value, and found similar results over the whole areas of the value except when $\mu$ approaches 1. That is why the dot-line ($\mu=0$, the structured model) and dash-line ($\mu=10^{-3}$, the growing model) show similar features, approaching not to +1 or -1 and some jumps. If a crossover is observed when $\mu$ approaches almost 1, it is from the clustered feature of the network than the one-dimensional structure \cite{klemm05}.

The \textit{de facto standard} is a well-known phenomenon in economics and business management \cite{arthur97}. De facto is Latin for `in fact' or `in practice'. A de facto standard is one that everyone seems to follow as an authorized standard, such as Microsoft Windows, Apple iPod, and VHS, a recording and playing standard for video cassette recorders. \textit{Lock-in}, a situation whereby a customer depends on a vendor for products and services and so difficult to move to another vendor without switching costs, is also well known \cite{arthur97}. When the network is dominated by the majorities, the majorities can be regarded as the de facto standard. Also, the convergence of log return to +1 or -1 as lock-in.  

The structured and growing models have several clusters, which leads to some different aspects, comparing with the BA and WS models. In those models, clustering of agents with the minority opinion can induce and maintain their own market share. At least, the minorities of those models can have a chance to be the majorities or expand their market share even though the possibility is low, but not the BA and WS models. When the market is dominated by one opinion, those who hold another opinion (or products or services) cannot expand their market share. The distinguished feature of the structured and growing models from others is the existence of high clusters. This can be applied to marketing strategies in practice. For example, the success of the Apple iPod began from a few enthusiasts \cite{kahney05}. Of course, clustering cannot guarantee success, but it is not easy to reverse the majority opinion without a high degree of clustering of customers, especially if there is already a de facto standard or lock-in.

\section{Discussions and conclusions}
We analyzed opinion dynamics in complex networks. First we investigated a simple tug-of-war game to identify the slow pace of fast change. The result of this game shows that a great number of information sources leads to hesitation in decision-making and slow change for the system.  If a node in a given network has many neighbors, this node is subject to a great influx of information. The density of these nodes leads to clustering in the network, with clusters maintaining their particular opinions against external information. We examined this property using the microscopic spin model as the opinion model in the Watts--Strogatz small-world and Barab\'asi--Albert scale-free network models.  In these complex networks, information diffuses through the whole network very rapidly. Most nodes in the network have the same opinions, which can be explained by a de facto standard, which everyone seems to follow as an authorized standard in the market, and a lock-in effect, whereby a customer depends on a vendor for products and services and finds it very difficult to move to another vendor. However, the result for the structured and growing scale-free network model with a high degree of clustering shows obvious differences. The clustering may be regarded as groups of enthusiasts who make their own market share against the majority opinion of the market, the de facto standard or the lock-in.

\begin{acknowledgements}
We are grateful to Hang-Hyun Jo, Konstantin Klemm, Sungho Han, Fengzhong Wang and Shlomo Havlin for fruitful discussions and helpful comments.
\end{acknowledgements}

\end{document}